# Probabilistic Photonic Computing

*Frank Brückerhoff-Plückelmann[1,2,3], Anna P. Ovvyan[1,2], Akhil Varri[1], Hendrik Borras[4], Bernhard Klein[4], C. David Wright[5], Harish Bhaskaran[6], Ghazi Sarwat Syed[3], Abu Sebastian[3], Holger Fröning[4], Wolfram Pernice[1,2]\**

[1]Physical Institute, University of Münster; Münster, 48149, Germany.
[2]Kirchhoff-Institute for Physics, University of Heidelberg; Heidelberg, 69120, Germany
[3]IBM Research Europe, Säumerstrasse 4, 8803 Rüschlikon, Switzerland.
[4]Institute of Computer Engineering, University of Heidelberg; Heidelberg, 69120, Germany.
[5]Department of Engineering, University of Exeter; Exeter, EX44QF, UK.
[6]Department of Materials, University of Oxford; Oxford, OX43PJ, UK.

\*Correspondence to: wolfram.pernice@kip.uni-heidelberg.de

**Probabilistic computing excels in approximating combinatorial problems and modelling uncertainty. However, using conventional deterministic hardware for probabilistic models is challenging: (pseudo) random number generation introduces computational overhead and additional data shuffling, which is particularly detrimental for safety-critical applications requiring low latency such as autonomous driving. Therefore, there is a pressing need for innovative probabilistic computing architectures that achieve low latencies with reasonable energy consumption. Physical computing offers a promising solution, as these systems do not rely on an abstract deterministic representation of data but directly encode the information in physical quantities. Therefore, they can be seamlessly integrated with physical entropy sources, enabling inherent probabilistic architectures. Photonic computing is a prominent variant due to the large available bandwidth, several orthogonal degrees of freedom for data encoding and optimal properties for in-memory computing and parallel data transfer. Here, we highlight key developments in physical photonic computing and photonic random number generation. We provide insights into the realization of probabilistic photonic processors and lend our perspective on their impact on AI systems and future challenges.**

**Keywords:** Probabilistic Computing, Neuromorphic Computing, Photonic Processors

# Introduction

The use of artificial neural networks (ANNs) has been steadily increasing in various application domains including autonomous driving [1], medical diagnosis [2], and natural language processing [3]. As utilization of these applications gains popularity, so does the demand for energy-efficient, low-latency hardware capable of handling the intensive computations required by ANNs. Particularly in safety-critical applications[4], there is an immediate need for reliable uncertainty estimation, especially for scenarios outside the training distribution. Traditional hardware solutions, such as CPUs, GPUs, and tensor processing units (TPUs), are inherently deterministic, making them ill-suited for evaluating probabilistic models without incurring significant computational overhead and increased latency.

Physical neuromorphic computing approaches emulate the working principles of biological brains by directly encoding the data in physical quantities instead of abstract representations of data. In particular, neuromorphic architectures leverage massively parallel in-memory computation which offers a promising route to reducing the latency and power consumption of AI accelerators[5]. Moreover, neuromorphic systems are designed to incorporate tunable stochasticity, aligning with the neuroscience principle of free energy minimization (FEM) [6,7]. This built-in stochasticity is essential for effective uncertainty estimation and solution approximation. As computation is implemented via manipulating physical quantities in an analog way, physical entropy sources for true random number generation can be directly linked to the computation without inducing further overhead. For example, thermal noise in a magnetic tunnel junction can be exploited to realize a probabilistic matrix weight based on a tunable probabilistic conductance distribution [8].

Photonic physical computation systems are especially intriguing due to the large available bandwidth, e.g., more than 4 THz only considering the default communication C-Band, and multiple orthogonal degrees of freedom to encode and manipulate data in parallel. This fueled the development of photonic hardware accelerators leveraging amplitude [9] and phase encoding [10] and orthogonal encodings like time-frequency in combination with dispersive fibers [11] or amplitude-frequency and amplitude-mode with the corresponding multiplexers [12,13]. In addition to the variety of physical quantities for data encoding and computation, photonics also offers optimal properties for data transmission and storage as no charging of wires is present and higher long-time memory stability for example in the case of phase change memory [9]. Similarly, photonics offers multiple ways for true random number generation, each linked to different physical quantities like phase and intensity [14,15] thus enabling a high compatibility with the various encoding and computation schemes.

Here we offer our viewpoint on recent advancements in physical probabilistic computing, with emphasis on photonic implementations. We examine current architectures from both hardware and software perspectives and highlight their fundamental working principles. Our focus is on the seamless integration of entropy sources with high-speed photonic processors and their utilization on a system level. Additionally, we offer an outlook on the impact of photonic probabilistic processors on artificial intelligence and discuss the challenges lying ahead.

## 1. Beyond Conventional Deterministic Computing

Specialized digital deterministic processors such as TPUs power state-of-the-art neural networks and keep pushing electronic hardware to its physical limits to deal with the massive computational workload at reasonable energy levels and time scales[16]. Emerging analog computing architectures promise to complement conventional systems for dedicated tasks such as massively parallel, low latency matrix multiplication[5,11,17]. In order to harness the advantages of probabilistic computing and design networks that incorporate stochasticity as one of the cornerstones of neuromorphic design, novel processors are required[18]. Since physical entropy sources lie at the heart of probabilistic sampling, their direct integration with physical computing architectures is especially intriguing.

### 1.1 Physical Computing

Physical processors offer tremendous potential to augment conventional hardware due to their fundamentally different architecture as illustrated in **Fig. 1**. In contrast to the abstract data encoding in digital processors, physical processors encode the information directly in a physical quantity, for example the phase, amplitude or length of a pulse [5,9,10,19]. Similarly, computation is directly implemented by manipulating the corresponding quantity, for example changing the amplitude of a pulse by a fixed factor to realize a multiplication. In contrast, digital processors use complex concatenations of logic gates to realize the same operation on the abstract digital data encoding. This has a severe impact on the overall properties of the system. Physical systems are inherently prone to noise, with shot noise arising from the discrete nature of photons and electrons imposing a lower limit and excess noise being present in practical systems, for example due to spontaneous emission of photons or Johnson–Nyquist noise in electronic circuits. Thus, physical processors, which directly encode the information in a physical quantity will always exhibit noise and limited precision. In contrast, the abstract (binary) encoding in digital systems effectively bypasses this restriction. Even though the state of a single bit is still linked to a physical quantity, e.g., the voltage, it's a binary system. Therefore, it's not affected as long as the physical noise cannot cause a bit flip that is not detected by parity checks, thus enabling high precision computing. This fundamental difference explains why conventional digital processors are ill-suited for probabilistic computing and sampling, even though there are efforts to realize probabilistic bits[20], whereas it is an inherent property of physical processors. The challenge and at the same time the unique opportunity lies within making the physical noise accessible and tunable such that it can be used for probabilistic modeling.

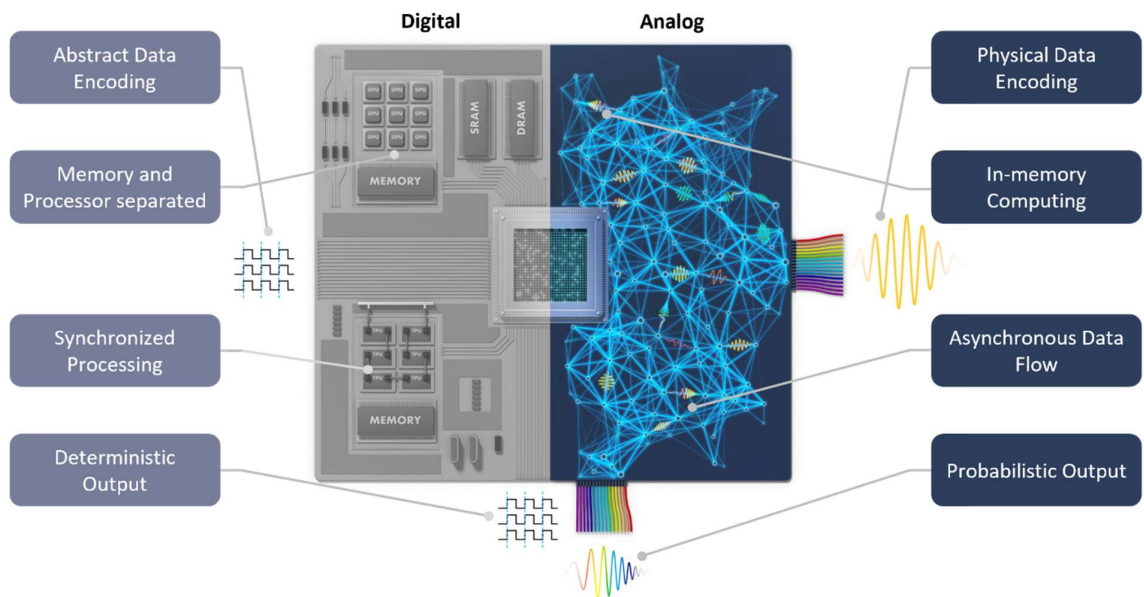

**Figure 1. Route towards probabilistic computing.** Probabilistic computing on digital hardware leads to significant overhead as the abstract data encoding is intrinsically designed for deterministic high precision computing. In contrast, physical computing is inherently probabilistic as the information is directly encoded in a physical quantity and computation is performed by manipulating those quantities. By making the innate noise accessible and tunable, probabilistic computing is possible without any computational overhead. In addition, the continuous evolution in time of the physical processor favors parallel low latency computation in comparison to the discrete clock based digital operation scheme.

In addition to the fundamental difference to digital computing regarding data representation, physical computing also inherently paves the way to massively parallel low-latency computation. Neither data movement nor computation is linked to a clock rate, instead the information carrier, e.g., the optical pulse, propagates through the physical system that implements the required mathematical operations, i.e. manipulations of the physical quantity encoded on the information carrier. As several information carriers (must) propagate through the physical system at the same time and the propagation time is in general short in comparison to the clock rate of the interface, low-latency parallel computation is an inherent feature. In contrast, a digital system would perform the same operations on subsequent time bins given by the clock rate, thus resulting in a significantly larger latency.

## 1.2 Photonic Architectures

The bosonic nature of photons makes them ideal information carriers in combination with low-loss and broadband optical interconnects. In addition, light pulses propagate in vacuum with around 300000 km/s and are only slowed down by a factor of two in typical waveguide materials as silicon nitride, reducing the inherent latency of a physical processor. As mentioned earlier, photonics offers a variety of physical quantities that can be used to encode data, including frequency [11,21], power [9,22], phase [10], polarization modes[12] and spatial modes [23]. Similarly, non-linear material responses may be used to manipulate the frequency of a pulse [24], tunable absorber and phase shifter can change the power and phase of a pulse [25,26] and tunable converters and multiplexers can change the mode states [27,28]. The combination of both leads to a variety of integrated photonic processors including waveguide crossbar arrays [9,29], Mach-Zehnder Interferometer meshes [10,30], microring weight banks [22,31] and diffractive networks[32,33]. While there exists an exhaustive number of reviews explaining the different computation schemes[17,34], it is important to consider the fundamental constraints and advantages of photonic processors to lend a perspective on their utilization for probabilistic computing. In general, the width of dielectric waveguides is in the order of the effective wavelength inside the medium, i.e. on a 0.4 μm to 1 μm scale considering the default C-Band [35]. As guiding is based on total internal reflection, smooth bend radii are required for routing, where the minimal bend radius depends on the refractive index contrast and might range between 5 μm for the Silicon-on-Insulator (SOI) platform and 30 μm for the silicon nitride on insulator (SiN) platform. Active components like laser sources, modulators and photodetectors also tend to be on a 100 μm scale. While plasmonic waveguiding can strongly reduce the feature size, ohmic losses drastically increase the propagation loss and cancel out one of the main advantages of photonic waveguides[36]. Guiding light in an integrated dielectric waveguide is possible with propagation losses as low as 1.77 dB/m [37], making large physical circuit designs feasible, which pairs well with the large feature sizes of photonic components that are compatible with larger fabrication nodes, e.g., 180 nm processes. A major challenge, however, is the large size of photonic components. Physical photonic computing would need to overcome significant challenges to closely pair with memory densities of physical electronic processors, i.e. having a matrix cell well below 1 μm². On the other hand, photonic is ideal for large circuit designs due to optimal propagation properties and coarser feature sizes. Thus, complex and compute heavy tasks that do not require a large amount of memory are an ideal fit for photonic processors. Overall, photonics is a compelling platform for physical computing, driven by multiple degrees of freedom for information encoding and manipulation in combination with the optimal waveguiding properties. These attributes collectively contribute to a reduced latency and minimal power consumption, making photonics an interesting approach for advanced computational hardware.

## 2. Photonic Entropy Sources

Entropy sources lie at the heart of probabilistic computing. Independent of the type of physical computing, there are two approaches for direct integration. First, integration within the computation units [8,38,39], second, as an innate feature of the physical information carrier [40–42]. While the first approach offers more flexibility for modeling the stochasticity, it strongly increases the complexity and potentially the energy dissipation of the circuit. In addition, random fluctuations in material properties might be bandwidth limited. As the power spectral density of a random process is directly linked to its autocorrelation function via the Wiener–Khinchin theorem, a small bandwidth $\Delta v$ leads to a large correlation time $\tau_c \sim 1/\Delta v$. Since subsequent samples must be uncorrelated for many applications, the correlation time induces an upper limit of the achievable sampling rate. Photonics offers various possibilities for innate random physical information carriers, including phase and intensity noise of classical light sources and quantum fluctuations on the single photon level. **Fig. 2** sketches the mechanism of the different entropy sources and how they can be leveraged for true random number generation.

### 2.1 Phase Noise

Pure phase noise manifests in a finite coherence time of an optical source while the second order of coherence, given by the power fluctuations, remains unity for all time delays. Practically, phase noise is present in laser systems and is primarily caused by spontaneous emission within the cavity in two ways. First, spontaneous emission can directly change the phase of the outgoing laser beam, second spontaneous emission can deplete the excited state of the gain medium and thus randomly change the refractive index along the laser cavity[43]. The second effect is mostly present in semiconductor lasers[44]. As phase noise at optical frequencies around 200 THz is not directly measurable via an intensity measurement, interferometric measurement schemes are deployed as illustrated in **Fig. 2a**. By superimposing a single laser with a delayed copy of itself (homodyne detection) or utilizing the difference between the phase of two separate lasers (heterodyne detection), physical random number generation can be realized [15,45–49]. While true random number generators based on photonic phase noise have been successfully realized, the deployment within computation schemes have been elusive up until now. Naturally, physical computation schemes deploying the phase as an information carrier, for example MZI-meshes[10] would be a suitable fit.

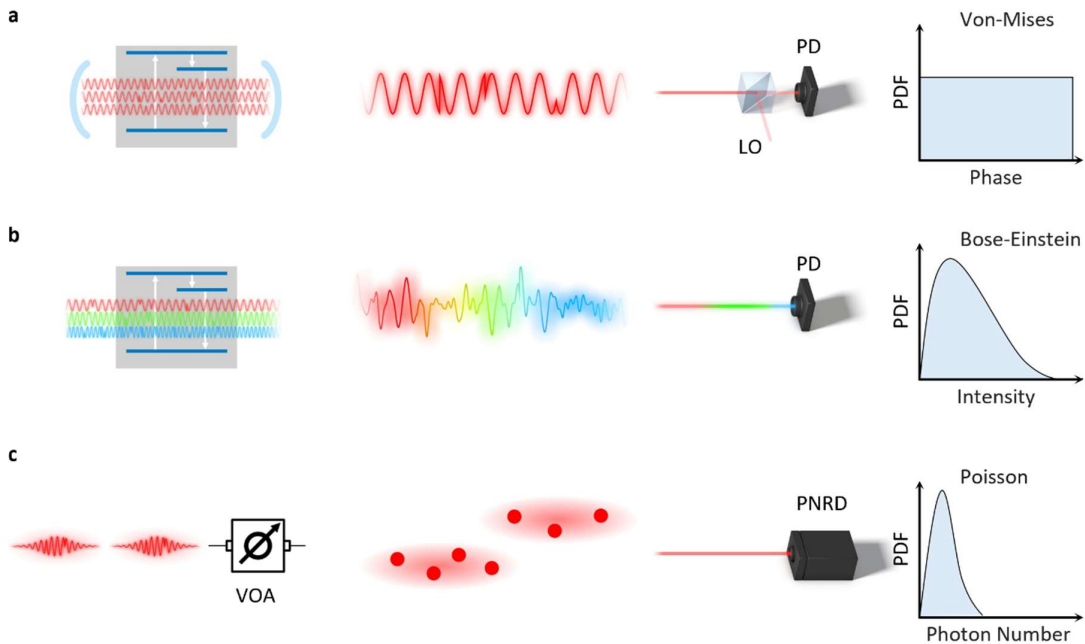

**Figure 2. Optical random number generation. a**, Spontaneous emission within laser cavities induces phase noise in the outgoing coherent laser beam defining its linewidth. The phase noise can be extracted via homodyne or heterodyne detection schemes and follows a von-Mises distribution. The distribution is uniform for mixing with an uncorrelated local oscillator (LO). **b**, Chaotic light states emitted by thermal light sources or generated by amplified spontaneous emission inherently feature power fluctuations due to the random beating between the various frequency components of the light state. The fluctuations can be directly measured with a photodetector (PD) and follows an M-fold Bose-Einstein distribution. **c**, Quantum fluctuations can serve as a perfect source of randomness on a single photon level, for example if the light state is not an eigenstate of the measurement operator. Thus, an optical random number generator might be directly implemented by generating (weakly) coherent state with a (pulsed) laser and a variable

optical attenuator (VOA) and measuring the state with a photon number resolving detector (PNRD). Here, the photon number distribution follows the Poisson statistics.

## 2.2 Intensity Noise

In contrast to phase noise, intensity noise leads to a non-constant second order of coherence, which describes the autocorrelation of the intensity for classical fields, in addition to a finite coherence time and is directly measurable with a photodetector. Intensity noise can be easily generated over a large bandwidth by exploiting photon bunching in chaotic light sources, for example amplified spontaneous emission in erbium doped fibers, laser diodes with the feedback [50], superluminescent light emitting diodes or thermal light sources. The intensity noise arises from the chaotic beating between the various frequency components of the chaotic light state resulting in a M-fold Bose-Einstein distribution, as sketched in **Fig. 2b**. The degeneracy factor M depends on the number of independent coherence cells within the measurement time and can be experimentally tuned by choosing the optical bandwidth of the chaotic carrier and the bandwidth of the electronic detection circuit[51]. An important property is that the standard deviation of the power fluctuations are proportional to the mean power, $\sigma \sim \mu/\sqrt{M}$, thus the standard deviation of the probability distribution can be modified at optical data transmission frequencies easily exceeding 10 GHz [42]. As the correlation time of the fluctuation is inversely proportional to the optical bandwidth, ultra high-speed physical random number generation can be realized[14,52–54]. There exist two approaches for direct integration into photonic processors. One is to build a mixed system using both coherent light states and chaotic states. In this way, the low-intensity noise of the coherent state may represent the mean of the distribution whereas the chaotic light state leads to noise around some mean, together enabling the modeling of distributions [40]. The drawback is the complexity of the physical system, as different states of light with different bandwidth need to be properly tuned with respect to each other and propagated through the system, practically limiting the scalability. Another approach would be to solely rely on chaotic light states, drastically simplifying the optical management in the circuit as chaotic light sources tend to be easier to realize as no cavity is required and the same light source can be deployed for each information instead of sufficiently detuned coherent lasers [42].

## 2.3 Quantum Fluctuations

Apart from exploiting the randomness of macroscopic systems manifested in the coherence properties of coherent and chaotic light at high photon numbers, the intrinsic randomness of quantum mechanics can be directly harnessed. For example, weak coherent states measured in the photon number basis exhibit a Poisson distribution, which can be exploited as a quantum random number generator as shown in **Fig. 2c**. Eaton et al., for instance, used transition-edge sensors as photon-number-resolving detectors to measure weak coherent states with photon numbers ranging from 0 to 100, thereby generating unbiased random numbers[55]. Similar experiments have been conducted using silicon photomultipliers, and these systems may eventually be fully integrated with photon-number-resolving superconducting nanowire single-photon detectors[56,57]. There are two approaches to leveraging these quantum fluctuations—i.e., the measurement of light states that are not eigenstates of the measurement operator—for computation. First, a processor might be designed to compute directly on quantum states, which leads to the field of photonic quantum computing, as for example demonstrated by L. Madsen et al. [58]. Second, classical light states could be coupled to quantum states, allowing the measured fluctuations to be tuned. In this case, the fluctuations would then be coupled to a macroscopic quantity, resulting in the measured probability distribution[59]. The work of Choi et al. characterizes the basic required components and proposes a potential design for a future all-optical system, although a functional prototype remains elusive [60].

## 3. Probabilistic Processors

Deterministic architectures augmented with (pseudo) random number generators are a powerful framework for various application fields including Bayesian neural networks (BNNs) [61,62], generative models [63,64] and heuristic methods [65]. Due to the deterministic nature of digital processing, pseudo RNG imposes a significant bottleneck that is detrimental to real-time tasks like confidence estimation during autonomous driving and might induce unreasonable convergence times for heuristic optimization. Inspired by the progress in analog, electric probabilistic computing for BNNs [38,66] and in-memory optimization [67,68], we lend our perspective on the utilization of integrated photonic probabilistic processors deploying chaotic light as the entropy source. Such processors concurrently allow for matrix-vector multiplication and massively parallel random number generation in a single clock cycle, providing a drastic speedup in comparison to pseudo random number generation and material-based entropy sources. **Fig 3a** sketches multiple methods to obtain tunable stochasticity in a photonic crossbar array solely based on power encoding. First, we can mix chaotic light with coherent light and thus control the mean and variance of the distributions by adapting the power ratio between them[40]. In order to reduce the complexity of the circuit, we can also solely utilize chaotic carriers[29]. However, this directly links the mean to the standard deviation of the distributions that are encoded, strongly limiting the flexibility. As the degeneracy factor of the underlying Bose-Einstein distribution depends on the number of independent coherence cells within the measurement interval, we need to access this quantity. Multi symbol waveform encodings can effectively change the measurement interval at GS/s rates by adapting the pulse length and height[42]. In contrast, changing the optical bandwidth will change the size of the coherence cells without requiring multi-symbol encodings. While this mechanism has not been realized yet, it promises the highest performance.

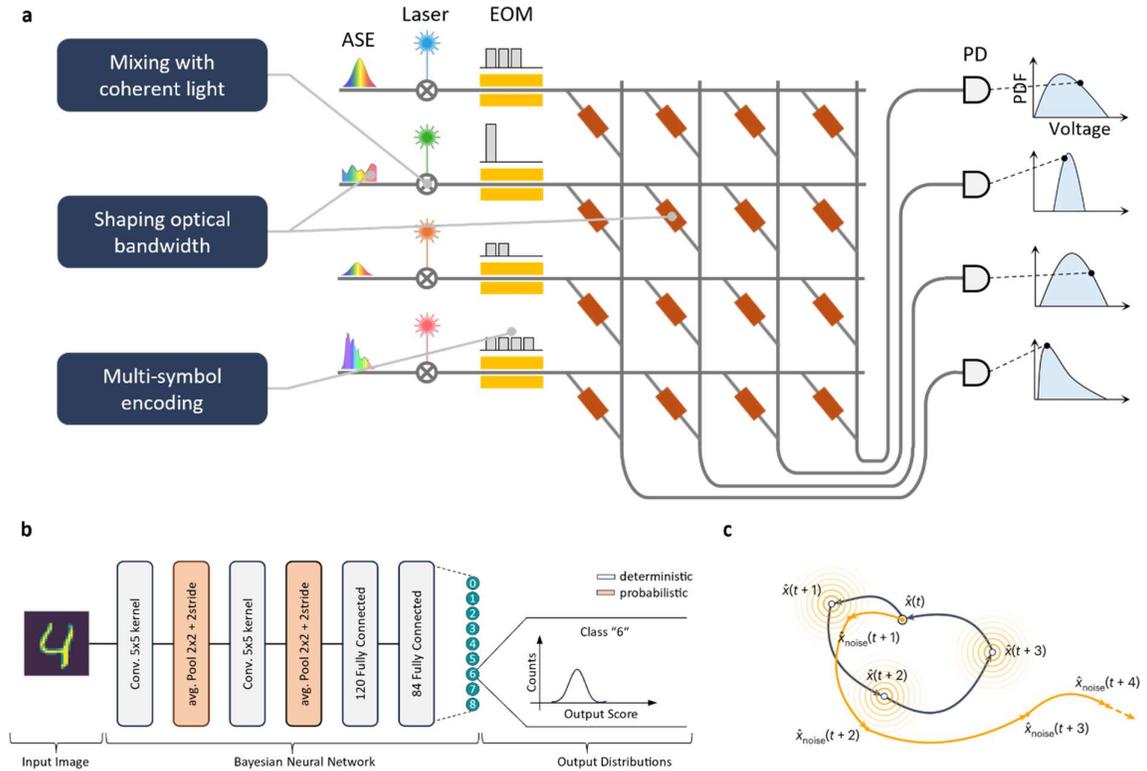

**Figure 3. Photonic probabilistic computing. a**, There exist several possibilities to enhance power-encoding based photonic crossbar arrays with tunable stochasticity. First, one can mix coherent carriers with chaotic carriers and use the power ratio to encode the mean and variance of the distributions. Second, one can only rely on chaotic carriers and modify the degeneracy factor of the underlying Bose-Einstein distributions. This has been demonstrated via multi-symbol encodings and is also feasible by adjusting the optical bandwidth within the circuit. The full probabilistic matrix vector multiplication (MVM) is implemented by a default optical transmission measurement. **b**, The probabilistic processor might be deployed to implement a probabilistic layer in a feedforward architecture like a Bayesian neural network. With a sub-ns latency for the MVM and sampling from the output distribution, photonics can efficiently remove the bottleneck

from probabilistic sampling arising in purely deterministic hardware. The picture is reproduced from[42] **c**, Injecting noise[68] in recurrent systems like Hopfield networks or vector symbolic architecture based factorizes greatly improves their performance as convergence to local minima can be avoided. Due to the sequential nature, these architectures also strongly benefit from the low-latency and highly parallelism of photonic processors. The picture is reproduced from[67].

## 4. Probabilistic Deep Neural Networks

Combining deep neural networks (DNNs) with random number generators for probabilistic sampling can greatly improve the training process, for example by augmenting the input data or enforcing a smoother latent space in encoder-decoder architectures, and enhance the performance of generative models. Probabilistic photonic computing, with its ability to simultaneously perform ultrafast random number generation and computation, offers significant potential to accelerate these networks.

### 4.1 Diffusion Models for Image Generation

Diffusion models[63] are widely used in generative modeling for producing high-fidelity images, such as those in DALL-E[69]. These models work by iteratively adding Gaussian noise to data and then learning to reverse the noise process to generate new samples. During inference, diffusion models require extensive sampling from noise distributions at each time step and passing these samples through a trained convolutional neural network, often a U-Net[70]. This sequential process makes image generation computationally intensive.

Probabilistic photonic computing can accelerate diffusion models by enabling ultrafast, high-quality random number generation, as shown in **Fig. 2**. Additionally, depending on matrix sizes and data flow, convolutional and pooling operations following sampling can be implemented optically at speeds of tens of GS/s. Recent demonstrations of optical non-linearity in silicon photonics [71] also enable sequential neural network computation directly within the optical domain, reducing unnecessary conversions between optics and electronics. This acceleration could enable real-time image generation and expand the use of diffusion models in time-sensitive applications such as video synthesis and interactive media.

### 4.2 Variational Autoencoders for Data Imputation

Variational Autoencoders (VAEs) [64] are a class of generative models that use a neural network (encoder) to transform high-dimensional input data into a lower-dimensional latent space, typically represented by probability distributions, such as Gaussian distributions. Samples are drawn from these distributions and passed through a decoder network to reconstruct the input data, enabling the generation of new samples. For example, genomic datasets often have missing or incomplete data due to limitations in sequencing technologies, sample quality, or cost constraints. VAEs can be applied to genomic data imputation by learning the underlying distribution of genetic variants[72].

Sampling from the latent space is a critical operation in VAEs. For genomic data imputation, VAEs can be computationally intensive due to the high dimensionality of the data and the need for extensive sampling from complex probability distributions. Probabilistic photonic computing could accelerate this process by speeding up the sampling operations inherent in VAEs and efficiently managing high-dimensional data across separate wavelength channels[42]. Like diffusion networks, implementing the neural network in this context poses challenges in synchronizing the hybrid optoelectronic approach to maximize the use of computing resources.

## 5. Bayesian Neural Networks

Bayesian neural networks (BNNs) fundamentally differ from other (probabilistic) DNNs in the sense that the network parameters are given by the distribution inferred from the known training data and not by a minimal cost point estimate[62]. Introducing probabilistic mappings within the architecture allows for additional degrees of freedom, for modelling uncertainty, as illustrated in **Fig. 3b**. However, the sequential operations and large state associated with pseudo-random number generation pose a significant bottleneck. For example, sampling rates for Mersenne Twister pseudo-random number generation are only on the order of 10 MS/s. Physical computing can remove this bottleneck, as it is inherently probabilistic. In the electronic domain, D. Bonnet et al. exploited the programming noise of memristors and phase-change materials within a crossbar array to implement a BNN [66]. To avoid the latency and potentially limited endurance associated with continuous reprogramming, S. Liu et al. utilized the thermal noise in magnetic tunnel junctions to enable probabilistic in-memory computing with a sampling rate of 500 MS/s [8]. In contrast, photonics offers the potential for sampling rates exceeding 10 GS/s, as the correlation time of the entropy source is not linked to material properties but to the coherence properties and bandwidth of the optical carrier, which easily exceeds 1 THz [42]. The drawback is that the memory density of integrated photonic processors is significantly smaller than that of their electronic counterparts, limiting the size of probabilistic mappings. Similarly, a large portion of the overall architecture likely needs to be deterministic to provide the required memory capacity, with only a few probabilistic mappings. This trend aligns well with ongoing research in Bayesian neural networks, which focuses on the number and placement of probabilistic layers within the architecture to maintain sufficient functionality[73].

## 5.1 Computer vision models for safe autonomous driving

Autonomous driving is a core application of modern AI systems, where computer vision models have become prevalent due to their practical advantages in terms of hardware and sensor requirements. Some manufacturers, like Tesla, have even developed entire autonomous driving systems that rely solely on camera input. For safety-critical applications, uncertainty estimation in predictions based on visual inputs is essential, as it enables the system to gauge confidence in its predictions. As a specific example, Kendall et al. demonstrate the application of uncertainty estimation in monocular depth regression, where incorporating uncertainty helps to assess prediction reliability[4]. This capacity helps to identify when additional caution or alternative actions are necessary, enhancing overall safety.

Probabilistic models further improve overall performance by explicitly accounting for uncertainty in the input data and detecting cases that fall outside the model's expertise, known as out-of-domain data. This approach provides real-time insights into the model's operational boundaries, allowing for safer and more reliable deployment. Probabilistic photonic computing can drastically reduce prediction latency, enabling a faster feedback loop that enhances both safety and decision-making. Additionally, photonic computing offers considerable improvements in energy efficiency, enabling more complex computer vision systems to operate within current power budgets.

## 5.2 Bayesian Models for Medical Imaging Support

High-resolution medical imaging, such as MRI and CT scans, plays a crucial role in guiding medical professionals' attention to potential anomalies or critical artifacts within the body. In this safety-critical domain, providing real-time diagnostic support is highly valuable for enabling interactive and precise diagnoses. However, delivering this level of responsiveness is challenging due to the immense data volume inherent to high-resolution images.

Probabilistic models, particularly BNNs, can enhance these diagnostic workflows by highlighting regions of interest with quantified uncertainty, helping doctors focus on potential abnormalities with greater confidence. By providing uncertainty estimates, BNNs allow physicians to understand the reliability of the model's predictions, aiding in decision-making[74]. Additionally, real-time feedback can be made feasible with photonic computing, which accelerates these complex models while maintaining high-resolution processing demands, ultimately supporting faster, more accurate, and safer diagnostic insights.

## 6. Energy-based Neural Networks and Resonator Networks

Energy-based neural networks and resonator networks reuse output states as input states, enabling a temporal evolution of the system. These architectures are particularly intriguing for photonic computing when the dominant computational workload involves (probabilistic) matrix-vector multiplications with constant matrix elements[75]. Hopfield networks and Boltzmann machines, which were awarded the Nobel Prize in Physics in 2024[76], highlight the compatibility with photonic computing[77,78]. Both are fully connected recurrent architectures that can be used, for example, as an associative memory or for optimization tasks. Depending on the application, network parameters can be trained using, for instance, the Hebbian learning rule for associative memory[79], or they may be directly derived from the energy function that describes the optimization problem[68]. Since the energy function of Hopfield networks and Boltzmann machines resembles that of the Ising model, a variety of problems, such as the traveling salesman or max-cut problem, can be approximately solved using these recurrent architectures[80–83]. Hopfield networks evolve deterministically, posing a high risk of converging to local minima in the energy landscape. In contrast, the probabilistic Boltzmann machine can escape these minima[83,84]. **Fig. 3c**, reproduced from the work of J. Langenegger et al.[67], illustrates the temporal evolution in a deterministic and probabilistic architecture. In addition to classical Hopfield networks and Boltzmann machines, emerging vector symbolic architectures and resonator networks are an intriguing application for physical probabilistic computing as MVMs are the dominant computational workload[67]. These architectures use quasi-orthogonal vectors, also known as code vectors, to encode different symbols. The quasi-orthogonality enables parallel computation on the superposition of different symbols, which greatly improves the performance for high-dimensional factorization[85,86]. Moreover, neuro-symbolic AI, which combines standard ANNs and vector symbolic architectures, allow for abstract reasoning and thus set new state-of-the-art performance in fluid intelligence tests like Raven's Progressive Matrices[87].

Probabilistic recurrent architectures impose two major computational challenges. First, their operation is sequential, meaning that latency must be minimized, in contrast to maximizing throughput in parallel architectures. Second, probabilistic sampling is required at each timestep. Physical in-memory computing is ideal for reducing latency since it is only limited by clock speed at the input/output interface and during postprocessing, but not at the computation stage. R. Khaddam-Aljameh et al. demonstrated MVMs with a 256 x 256 matrix within 127 ns on an electronic crossbar array[88], while photonics holds promise for sub-nanosecond latency due to higher bandwidth and ideal memory properties that enable pulse amplitude modulation instead of pulse width modulation. First prototypes, like the Photonic Arithmetic Computing Engine (PACE) from Lightelligence, claim 150 ps optical latency for a 64 x 64 MVM and a latency of 3 ns for a single probabilistic iteration of the corresponding Ising problem[89]. Integrating a photonic entropy source directly into processors, as depicted in **Fig. 3a**, eliminates the need for electronic probabilistic sampling and even enables stochasticity that is tunable during the temporal evolution of the system.

## 7. Future Trajectory of Probabilistic Photonic Processors

Since noise is inherent to all physical processes that underly computation, an exciting direction is to explore options that do not suppress noise, but rather leverage it as a source of stochasticity. For instance, noise can be interpreted as random variability that facilitates stochastic sampling, potentially reducing the need for complex and computationally expensive sampling techniques typically required in purely digital implementations. By reconceptualizing noise as a resource rather than an error, photonic analog computing may open new pathways for applications that rely on probabilistic reasoning, particularly in energy-constrained environments where digital sampling methods may be impractical.

However, developing photonic entropy sources and processors from research demonstrators to proof-of-concept prototypes that can have an impact on nowadays computational challenges require a sophisticated design flow as sketched in **Fig. 4.** Defining application requirements is the foundational element that shapes the entire design flow. Key factors include the type of problem in terms of network architecture, availability of training data, enforced reliability standards, and inference performance targets, such as throughput, latency, and efficiency - all of which will influence the overall design.

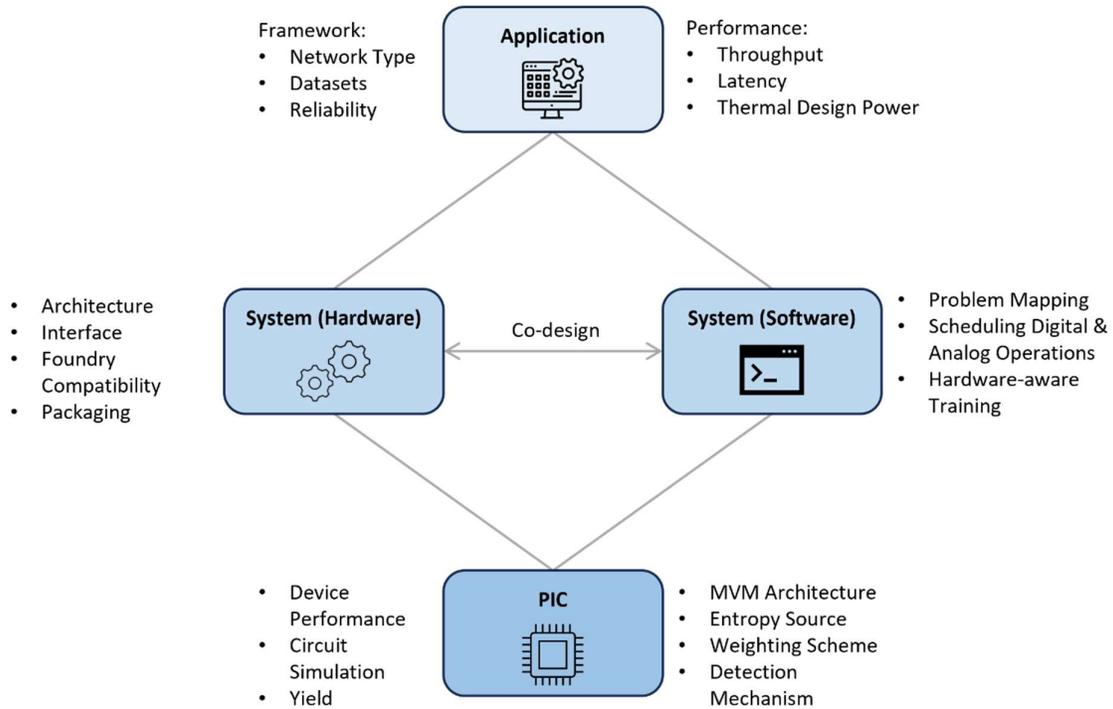

**Figure 4. Hardware-software co-design.** Developing probabilistic photonic processors from research experiments to proof-of-concept prototypes requires a sophisticated design strategy. First, the top application layer defines system requirements in terms of performance metrics and the general framework. The middle layer involves the interdependent development of hardware and software. Co-design ensures that the hardware aligns with software requirements, while the software accounts for hardware limitations and imperfections. Finally, the system architecture determines the design parameters of the photonic integrated circuit. Based on computing needs, such as memory density and noise encoding flexibility, the appropriate MVM architecture and photonic implementation are selected.

From a system design perspective, hardware non-idealities, such as fabrication imperfections, can negatively impact the performance of analog photonic circuits. Therefore, the software must account for these uncertainties during training. Hardware-aware training, which incorporates simulation or experimental data from devices, is gaining traction as a robust solution[90]. In-situ training[91] is another promising approach, where backpropagated gradients are experimentally calculated on-chip, inherently accounting for hardware noise during training.

Moreover, control algorithms must be implemented in software to manage temperature fluctuations in sensitive optical components like ring resonators. For instance, algorithms such as proportional-integral-derivative (PID) control can be implemented more compactly on-chip, using photodetectors as integrators, requiring careful consideration before chip tapeout. Peripheral hardware choices, including the laser source, receiver, and data converters, will also affect the system's effective number of bits (ENOB)[92]. Similarly, hardware-aware training based on techniques such as quantization-aware training or post-training quantization are required to maintain inference accuracy at lower ENOB.

Finally, the heterogeneous integration of optical sampling, photonic tensor cores, electronic tensor cores, and shared memory resources necessitates complex scheduling, partitioning, and pipelining during inference. These operations are often handled by off-chip peripherals, such as field-programmable gate arrays (FPGAs), limiting demonstrations to smaller scales. Large-scale in-memory computing architectures for DNN workloads can be realized by connecting multiple analog crossbar arrays, digital compute cores, and memory cores [93,94]. This architecture features a dense 2D mesh of on-chip interconnects for ultra-fast communication between cores, along with power management solutions and resource allocation for various DNNs across multiple chips. Similar architectures could be envisioned for photonic probabilistic computing, adding further complexity to the software-hardware co-design with the inclusion of sampling operations.

Overall, the combination of probabilistic and photonic computing presents promising avenues for enabling more secure and accurate Machine Learning algorithms, as well as realizing their deployment on fast and efficient hardware. While current examples are often at the lab-demonstration stage, the groundwork has been laid for future large-scale implementations. The interdisciplinary collaboration between photonic system architects and computer scientists in conjunction with industry partners for defining the application requirements and providing the required devices on a scalable foundry level will be crucial to enable the next generation of proof-of-concept prototypes.

## Acknowledgements

We thank Jochen Stuhrmann, from Illustrato, for his assistance with the illustrations. This research was supported by the European Union's Horizon 2020 research and innovation programme (grant no. 101017237, PHOENICS project) and the European Union's Innovation Council Pathfinder programme (grant no. 101046878, HYBRAIN project). We acknowledge funding support by the Deutsche Forschungsgemeinschaft (DFG, German Research Foundation) under Germany's Excellence Strategy EXC 2181/1 – 390900948 (the Heidelberg STRUCTURES Excellence Cluster), the Excellence Cluster 3D Matter Made to Order (EXC-2082/1—390761711) and CRC 1459 'Intelligent Matter'.

## Author contributions

- Conceptualization: FBP, WP
- Methodology: FBP, AO, AV, Hendrik B, BK, GSS
- Investigation: FBP, AO, AV, Hendrik B, BK, GSS
- Visualization: FBP, AO, AV
- Funding acquisition: WP, HF, CDW, Harish B, AS, HF
- Project administration: WP, HF
- Supervision: WP, HF, CDW, Harish B, GSS, AS
- Writing – original draft: FBP, AO, AV
- Writing – review & editing: All authors

## Competing interests

The authors declare that they have no competing interests.

## Additional information

Correspondence and requests for materials should be addressed to W.P.

## Data availability

All data is available in the main text.